# A Catalog of Exoplanets with Equilibrium Temperature ≤ 600 K

David G. Russell
*Owego Free Academy, Owego, NY USA*
Email: russelld@oacsd.org

*Abstract*

The NASA Exoplanet Archive was searched for planets with an equilibrium temperature ≤ 600 K, mass uncertainty ≤ 27%, and radius uncertainty ≤ 8%. Planets were included in the catalog if the mass was < 5.5 Jupiter masses. This search produced 93 planets with mass from 0.3 to 1680 $M_⊕$; and 101 planets if the Solar System planets are included. The main characteristics of the sample in this catalog are: (1) 94% of the Terrestrial planets have mass < 2.9 $M_⊕$ and radius < 1.4 $R_⊕$, (2) Most Neptune composition planets ($f_{H-He}$ 1.0 – 49.9% by mass) have a mass in the range 6.0 to 18 $M_⊕$ and a radius in the range 2.3 – 4.1 $R_⊕$, (3) The sample has a small drop in population consistent with the previously identified radius gap from 1.5 – 2.0 $R_⊕$, (4) Planets in the radius range 1.50 – 2.25 $R_⊕$ are consistent with either a gas-rich Terrestrial composition ($f_{H-He}$ 0.01 – 0.99% by mass) or a rock-ice Terrestrial composition with a supercritical hydrosphere and water mass fraction < 20%, (5) A super-Neptune radius desert is observed for the radius range 4.5 – 7.5 $R_⊕$, (6) Saturn composition planets ($f_{H-He}$ 50.0 – 80.0 % by mass) have masses from 15 – 170 $M_⊕$ and radii from 7.9 – 10.1 $R_⊕$, (7) A nearly barren **sub-Saturn mass-radius desert** is found in the sample as indicated by a lack of planets with mass exceeding 20 $M_⊕$ and radii in the range 4.0 – 7.5 $R_⊕$, (8) Most Jupiter composition planets ($f_{H-He}$ 80.1 – 100.0 % by mass) have radii between 10.9 and 12.4 $R_⊕$ and mass exceeding 200 $M_⊕$, (9) With few exceptions, planet radius can be used as a proxy for planet composition classification into Terrestrial, gas-rich Terrestrial or supercritical hydrosphere Terrestrial, Rock-Ice Giant, and Gas Giant composition classes for this sample of Teq ≤ 600 K planets. The characteristics of this sample are consistent with several predictions of the core accretion model including the predicted values for the critical core mass for gas accretion, the critical core mass for runaway accretion, the pebble isolation mass, and the Saturn mass desert.

## ~1. Introduction

With over 5200 confirmed planets in the NASA Exoplanet Archive database (Akeson et al. 2013) it is now possible to acquire a useful sample size while employing specific selection criteria for characteristics such as planetary radius, mass, density, composition, orbital period, equilibrium temperature, data uncertainty, habitable zone, or stellar characteristics such as effective temperature or spectral class (e.g. Dai et al. 2019; Otegi et al. 2020; Millholland et al. 2020; Carríon-González et al. 2021; Cuntz et al. 2022; Hill et al. 2022).

Most confirmed planets in the NASA Exoplanet Archive have shorter orbital periods and higher stellar flux values than the planets of the Solar System. As a result, among the exoplanets that have similar mass and radius values for comparison with specific Solar System planets, few also have low equilibrium temperatures, i.e. less than 450 K. For example, TOI-178 b (Leleu et al. 2021) has a mass 50% larger than Earth and a composition that is ~20% Fe and 80% silicates from the models of Zeng et al. (2016). However, TOI-178 b has a high equilibrium temperature (Table 1) and is therefore consistent with a planet that may have an "Earth-like" composition but cannot have "Earth-like" conditions. The mass and radius values of TRAPPIST-1 c (Table 1) are consistent with a planetary composition that includes a 25% by mass Fe core and a 75% by mass silicate mantle (Zeng et al. 2016). The stellar flux of TRAPPIST-1 c is 2.1214 $S_⊕$ (Agol et al. 2021) and therefore this planet is one of the closest to "Earth-like" planets found in the NASA Exoplanet Archive.

There are several reasons a sample of planets with equilibrium temperatures less than 600 K can be useful. Planets with a low equilibrium temperature can more closely represent analogs for the conditions of Solar System planets. Earth-like composition planets with very high values of stellar irradiation may have molten surfaces and be "lava worlds" (Chao et al. 2021) that do not match the surface conditions of the Solar System's Terrestrial planets.

**Table 1: TRAPPIST-1 c and TOI-178 b planetary data**

|  | TRAPPIST-1 c | TOI-178 b |
|---|---|---|
| **Period (days)** | 2.42 | 1.91 |
| **Mass ($M_\oplus$)** | 1.308 +/- 0.056 | 1.50 +0.39/-0.44 |
| **Radius ($R_\oplus$)** | 1.097 +0.014/-0.012 | 1.152 +0.073/-0.070 |
| **Teq (K)** | 340 | 1040 |
| **Flux ($F_\oplus$)** | 2.1214 | 216 |
| **Composition** | 25% Fe core / 75% Silicates | 20% Fe core / 80% Silicates |
| **Reference** | Agol et al. 2021 | Leleu et al. 2021 |

Composition from Zeng et al. (2016) models.

Terrestrial planets with a Ganymede-like composition can potentially retain surface liquid water up to temperatures of ~650 K (Nixon & Madhusudhan 2021). Any such worlds are therefore stronger candidates for habitable environments at the base of the atmosphere than more highly irradiated planets. Photoevaporation of the H-He envelopes for planets with lower masses and lower H-He fractions is expected to be greater at higher equilibrium temperatures (Lopez & Fortney 2013). Many Terrestrial composition planets at high equilibrium temperatures are likely the exposed cores of sub-Neptune composition planets (Dai et al. 2019) rather than analogs for the Solar System's Terrestrial planets.

Planets with an H-He envelope can experience significant radius inflation as equilibrium temperature increases with increasing stellar flux (e.g. Fortney et al. 2007; Lopez & Fortney 2014; Zeng et al 2019). Therefore planets should provide closer analogs to the Solar System's Gas Giants and Rock-Ice Giants at lower equilibrium temperatures. Lopez & Fortney (2014) provided models for the radii of Neptune and sub-Neptune mass exoplanets with H-He fractions ranging from 0.01 to 20% by mass and for stellar flux values of 0.1, 10, and 1000 $S_\oplus$. These models indicate that increasing the stellar flux from 0.1 $S_\oplus$ (Teq ≈ 150 K) to 10 $S_\oplus$ (Teq ≈ 500 K) results in modest radius inflation of 1-7% for planets in the mass range 5.5 to 20 $M_\oplus$ and with 0.5 to 20% by mass H-He. Increasing the stellar flux from 0.1 to 1000 $S_\oplus$ (Teq ≈ 1570 K) causes significantly larger radius inflation ranging from 8 – 40%.

Radius inflation resulting from high stellar flux is also expected to occur within the population of Gas Giant planets (Fortney et al. 2007). For example, the Jupiter mass planet Kepler 167e ($M_\oplus$ = 321) has a low equilibrium temperature of 134 K and a radius of 10.16 $R_\oplus$ (Chachan et al. 2022), within 8% of Jupiter's radius (10.97 $R_\oplus$). In contrast to Kepler 167e, the Jupiter mass planet TOI-1601 b ($M_\oplus$ = 315) has an equilibrium temperature of 1619 K and a radius of 13.89 $R_\oplus$ (Rodriguez et al. 2021), which is 26.6% larger than Jupiter. Alves et al. (2022) found that the 750 $M_\oplus$ super-Jupiter NGTS-21 b, with a stellar flux of 1000 $S_\oplus$, is inflated ~21% ($R_\oplus$ = 14.6).

Millholland et al. (2020) found that planets with H-He envelopes can also have radii inflated from tidal effects. After accounting for tidal inflation in a sample of sub-Saturn radius planets, Millholland et al. (2020) found planets had lower H-He fractions than estimated from models that do not account for tidal inflation, such as Lopez & Fortney (2014). Planets in the Millholland et al. (2020) sample with orbital periods less than 12 days had H-He mass fractions only 25-50% of the values found from non-Tidal models whereas planets with orbital periods over 25 days experience significantly smaller tidal inflation. The results of Millholland et al. (2020) add another factor that will tend to make short orbital period giant planets with high Teq less suitable analogs for comparison with the giant planets of the Solar System.

In this paper a catalog of planets identified from the NASA Exoplanet Archive (Akeson et al. 2013) with equilibrium temperatures ≤ 600K, uncertainty in radius ≤ 8%, and uncertainty in mass ≤ 27% is presented. In section 2 the sample selection criteria is described. In section 3, the general characteristics of the composition classes are overviewed. In section 4 the catalog population characteristics are considered in the context of previous findings from the literature and a number of characteristics of the population relevant to formation mechanisms such as core accretion are noted. Section 5 is a summary of the main characteristics of the catalog sample.

## ~2. Sample Selection

The NASA Exoplanet Archive was searched for planets meeting the following criteria:

~The planet has an equilibrium temperature ≤ 600 K within the +/- uncertainty of the equilibrium temperature (Teq), or the planet has a stellar flux ($S_⊕$) ≤ 25.0 within the +/- uncertainty of the stellar flux value.

~The planet has a mass value with uncertainty ≤ 27%. Planets were included in the catalog sample up to a mass of 1750 $M_⊕$ or 5.5 Jupiter masses.

~The planet has a radius value from the same source as the mass value with uncertainty in the radius ≤ 8%. An exception was made for Kepler 10 c, which has an independent reanalysis of the planetary mass that resulted in a significantly smaller mass than found in prior studies (Rajpaul et al. 2017).

The mass and radius uncertainty limits are similar to those used by Otegi et al. (2020). When available, planetary Teq or $S_⊕$ was taken from the same source as the mass and radius values. If no Teq or $S_⊕$ was provided by the mass and radius data source, then the stellar flux from Berger et al. (2018) was used if available. Several planets were included in the sample that have no available Teq or $S_⊕$ in the NASA Exoplanet Archive if it was concluded the planet must have an equilibrium temperature < 600 K based upon (1) the planet having a greater orbital semi-major axis than another planet orbiting the same star that does have Teq < 600 K or (2) the planet having a larger orbital period in an orbit around a star with a similar effective temperature when compared with other planets in this sample that meet the Teq requirement. Only four planets in this catalog have no Teq or $S_⊕$ provided but meet one of these two requirements. In total, 71 of the 93 planets in the catalog have a Teq or $S_⊕$ from the same source as the mass and radius values.

Since many planets have multiple entries in the NASA Exoplanet Archive database, the entry included in the catalog was either the most recent entry or the entry with the lowest uncertainty in mass and radius – which in most cases is the most recent entry.

## ~3. The Catalog

The 93 planets with Teq ≤ 600 K in this catalog are grouped into Tables 3, 4, 5, 6, and 7 based upon the composition classes described in Table 2, which includes both general composition classes and bulk composition classes characterized with Solar System analog names (Russell 2021). The general composition class of each planet is determined using the mass, radius, and Teq or $S_⊕$ values which are compared with mass-radius-composition models (Fortney et al. 2007; Lopez & Fortney 2014; Zeng et al. 2016, 2019; Aguichine et al. 2021; Vivien et al. 2022). The general composition classes ***Rock Terrestrial***, ***Rock-Ice Terrestrial***, ***gas-rich Terrestrial***, ***Rock-Ice Giant***, and ***Gas Giant*** are based upon the mass fractions of H-He gas, Fe and silicate rock, and $H_2O$ (Russell 2021).

### ~3.1 Rock and Rock-Ice Terrestrial Composition Planets

Rock and Rock-Ice Terrestrial planets have $f_{H-He}$ < 0.01% by mass and are listed in Table 3. The possible bulk compositions for these planets and corresponding Solar System analog names depend upon the Fe-silicate-water fractions (Table 2). Mercury and Earth composition Terrestrial planets have $f_{H2O}$ < 0.2% by mass whereas Europa and Ganymede composition Terrestrial planets have $f_{H2O}$ > 0.2% and >10% by mass respectively (Russell 2021). Model degeneracy in the Zeng et al. (2016) models allows for planets with mass and radius consistent with a pure Fe-silicate composition, but low Fe fraction (< 15 % by mass) to have a higher Fe fraction with a non-negligible $H_2O$ fraction. For this reason, Terrestrial planets are not identified as Mercury, Earth, Europa, or Ganymede composition in Table 3. All planets in Table 3 are consistent with $f_{H-He}$ < 0.01% by mass from the models of Lopez & Fortney (2014), and all but one have radii < 1.40 $R_⊕$.

**Table 2: General and Bulk Planetary Composition Classes**

| Composition | Mass (M⊕) | $f_{H-He}$ (%) | Notes |
|---|---|---|---|
| *Terrestrial* | | *<0.01* | |
| **Mercury** | < 3.0 | <0.01 | ~ $f_{Fe}$ > 50%, $f_{silicates}$ < 50%, $f_{H2O}$ < 0.2% |
| **Earth** | < 3.0 | <0.01 | ~ $f_{Fe}$ < 50%, $f_{silicates}$ > 50%, $f_{H2O}$ < 0.2% |
| **Europa** | < 3.0 | <0.01 | ~ $f_{Fe/silicates}$ 90 - 99.8%, $f_{H2O}$ 0.2 - 10% |
| **Ganymede** | < 3.0 | <0.01 | ~ $f_{Fe/silicates}$ < 90%, $f_{H2O}$ > 10% |
| **Super-Mercury** | 3.0 – 10 | <0.01 | ~ $f_{Fe}$ > 50%, $f_{silicates}$ < 50%, $f_{H2O}$ < 0.2% |
| **Super-Earth** | 3.0 – 10 | <0.01 | ~ $f_{Fe}$ < 50%, $f_{silicates}$ > 50%, $f_{H2O}$ < 0.2% |
| **Super-Europa** | 3.0 – 10 | <0.01 | ~ $f_{Fe/silicates}$ 90 - 99.8%, $f_{H2O}$ 0.2 - 10% |
| **Super-Ganymede** | 3.0 – 10 | <0.01 | ~ $f_{Fe/silicates}$ < 90%, $f_{H2O}$ > 10% |
| | | | |
| *Gas-Rich Terrestrial* | | *0.01 – 0.99* | |
| **Gas-rich Earth** | < 3.0 | 0.01 – 0.99 | |
| **Gas-rich Europa** | <3.0 | 0.01 – 0.99 | Hycean planets |
| **Gas-rich Ganymede** | < 3.0 | 0.01 – 0.99 | Hycean planets |
| **Gas-rich super-Earth** | 3.0 – 30 | 0.01 – 0.99 | |
| **Gas-rich super-Europa** | 3.0 – 30 | 0.01 – 0.99 | Hycean planets |
| **Gas-rich super-Ganymede** | 3.0 – 30 | 0.01 – 0.99 | Hycean planets |
| | | | |
| *Rock-Ice Giants* | | *1.0 – 49.9* | |
| **Gas-poor sub-Neptune** | 2.5 – 10.0 | 1.0 – 4.0 | |
| **Gas-poor Neptune** | 10.0 – 30.0 | 1.0 – 4.0 | |
| **Gas-poor super-Neptune** | >30.0 | 1.0 – 4.0 | |
| **Sub-Neptune** | 2.5 – 10.0 | 4.0 – 20.0 | |
| **Neptune** | 10.0 – 30.0 | 4.0 – 20.0 | |
| **Super-Neptune** | >30.0 | 4.0 – 20.0 | |
| **Gas-rich sub-Neptune** | 2.5 – 10.0 | 20.0 – 49.9 | |
| **Gas-rich Neptune** | 10.0 – 30.0 | 20.0 – 49.9 | |
| **Gas-rich super-Neptune** | >30.0 | 20.0 – 49.9 | |
| | | | |
| *Gas Giants* | | *50.0 – 100.0* | |
| **Sub-Saturn** | 15 - 40 | 50.0 – 80.0 | |
| **Saturn** | 40 – 200 | 50.0 – 80.0 | |
| **Jupiter** | 130 - 1300 | 80.1 – 100.0 | |
| **Super-Jupiter** | >1300 | 80.1 – 100.0 | |

*~3.2 Gas-rich Terrestrial or Supercritical Hydrosphere Terrestrial Composition Planets*

As discussed in Russell (2021), "gas-rich Terrestrial" planets (Coleman & Nelson 2016; Owen et al. 2020; Bean et al. 2021) are a transitional composition class between the Terrestrial (super-Earth and super-Ganymede) and Rock-Ice Giant (Neptune) composition classes. There is no consensus as to where the transition between super-Earths and sub-Neptunes is located. Lopez & Fortney (2014) suggested that a 5 M⊕ planet with a 0.5% by mass H-He envelope should be classified as a sub-Neptune because it would have a surface pressure ~20 x higher than the Marianias Trench and surface temperature ≥3000 K. Lozovsky et al. (2018) adopted an H-He envelope fraction of 2% by mass as the transition between super-Earths and sub-Neptunes. Russell (2021) suggested an H-He fraction of 1% by mass can serve as the composition boundary that divides the gas-rich Terrestrial population from the sub-Neptune population. In this catalog, planets that have mass-radius-Teq values consistent with a $f_{H-He}$ from 0.01 – 0.99 % by mass are identified as gas-rich Terrestrial. Most of the radius for planets with gas-rich Terrestrial composition results from the Terrestrial Fe-silicate or Fe-silicate-H₂O layers (Lopez & Fortney 2014).

The gas-rich Terrestrial planets are listed in Table 4 and most have radii in the range 1.50 – 2.25 $R_⊕$. Due to model degeneracy it is also possible some of the planets in this mass-radius space could be Terrestrial planets with a high mean molecular weight envelope (e.g. Cherubim et al. 2022), or Rock-Ice Terrestrial planets with a supercritical hydrosphere (Mousis et al. 2020; Aguichine et al. 2021; Turbet et al. 2021; Vivien et al. 2022). If gas-rich Terrestrial composition, most of the planets in Table 4 are consistent with $f_{H-He}$ in the range 0.1 – 0.5 % by mass from the models of Lopez & Fortney (2014). If the planets in Table 4 have a supercritical hydrosphere, then the water fraction for most is significantly less than 20% based upon the models of Aguichine et al. (2021).

### ~3.3 Rock-Ice Giant Composition Planets

Planets often characterized as "sub-Neptunes", "Neptunes", and "super-Neptunes" can have very different composition from the Solar System's "Ice Giants". Models for Neptunes allow core rock and ice fractions ranging from a pure rock core to a mixed rock-ice core with a water fraction as large as ~75% by mass (Marcus et al. 2010). In addition, there is significant variation in the mass fraction of H-He gas among the sub-Neptune, Neptune, and super-Neptune mass planets (e.g. Lozovsky et al. 2018; Millholland et al. 2020). Models for Uranus and Neptune favor a water dominated interior with a 10 – 15% by mass H-He envelope, but the observed parameters do not exclude models with a rock dominated, rather than $H_2O$ dominated, interior (Teanby et al. 2019; Helled et al. 2020). Vazan et al. (2022) noted that rock and ice may be fully miscible in the interior of 5 – 15 $M_⊕$ planets. Therefore, Neptune composition planets may lack fully differentiated layers of rock and ice and instead have mixed rock-ice interiors with density gradients (Vazan et al. 2022).

Considering the above challenges to "Neptune" planet composition characterization, Russell (2021) suggested that the term "Rock-Ice Giant" is a more suitable general term for this class of planet than "Ice Giant". As indicated in Tables 2 and 8, the composition of Rock-Ice Giants in this catalog can be divided into three bulk composition classes based upon the mass fraction of H-He gas ($f_{H-He}$). The H-He fractions indicated are consistent with the radius ranges indicated for the $T_{eq}$ ≤ 600 K sample listed in Table 5 of this catalog:

~*Gas-poor Neptune* composition planets are 1 – 4 % by mass H-He and have radii from 2.25 – 3.0 $R_⊕$.

~*Neptune* composition planets are 4 – 20 % by mass H-He and have radii from 3.0 – 4.5 $R_⊕$.

~*Gas-rich Neptune* composition planets are 20 – 49 % by mass H-He and have radii from 4.5 – 7.5 $R_⊕$.

Planets with these Neptune composition classes can also be characterized with three mass ranges as sub-Neptunes (2.5 – 10 $M_⊕$), Neptunes (10 – 30 $M_⊕$), or super-Neptunes (> 30 $M_⊕$). When combining the three planetary mass ranges with the three Neptune composition classes there are nine Neptune composition planet characterizations (Table 2).

The sample of 33 Rock-Ice Giants with $T_{eq}$ ≤ 600 K (Table 5) includes only a single exception to the radius ranges above. This high correspondence between radius and composition may be expected as a result of the smaller impact of radius inflation from the lower stellar flux values (Fortney et al. 2007; Lopez & Fortney 2014) and smaller tidal effects (Millholland et al. 2020) when compared with more highly irradiated Neptunes. The single exception is the young "super-puff" planet Kepler 51 d with a radius of 9.46 $R_⊕$ and an estimated age of 500 million years (Libby-Roberts et al. 2020). The radius of Kepler 51 d is consistent with the radii of Saturn composition Gas Giants (Tables 6 and 8). However, modeling the evolution of Kepler 51 d indicates that, starting with an estimated 39% H-He by mass, Kepler 51 d will retain 34% by mass H-He and have a radius of 6.2 $R_⊕$ after 5 Gyr (Libby-Roberts et al. 2020). Therefore, after 5 Gyr, Kepler 51 d will be consistent in both H-He mass fraction and radius with the "gas-rich Neptune" composition class and will be similar in mass, radius, and H-He fraction to the "super-puff" Kepler 87 c (Ofir et al. 2014). With a mass of 5.70 $M_⊕$, Kepler 51 d can therefore be characterized as a "*gas-rich sub-Neptune*" (see Table 2).

### ~3.4 Gas Giant Composition Planets

Gas Giant planets are divided into two bulk composition classes identified as Saturns ($f_{H-He}$ 50.0 – 80.0 % by mass) and Jupiters ($f_{H-He}$ 80.1 – 100.0 % by mass). The radii for the Saturn composition planets is in the range 7.9 – 10.1 $R_\oplus$ (Table 6). For Jupiter composition planets, the sample radius range is 10.1 – 13.5 $R_\oplus$ (Table 7). It is important to note that there is almost no mass overlap between the Saturn and Jupiter composition planets in the catalog. With the exception of TOI-216 b ($M_\oplus$ = 168), all Saturn composition planets in this sample have a mass < 130 $M_\oplus$ whereas all Jupiter composition planets have a mass > 130 $M_\oplus$. It is not clear whether the limited mass range overlap between Saturn and Jupiter composition Gas Giants in this sample is an expected characteristic resulting from formation mechanisms and the smaller effects of radius inflation for the population of Gas Giants having Teq ≤ 600 K, or instead results from a failure of surveys conducted so far to have discovered example planets in a larger existing mass overlap range. Given that three of the eleven Saturn composition planets have a mass that overlaps the Neptune composition mass range from 15 – 30 $M_\oplus$, it is noteworthy that there is no overlap of Jupiter composition planets with the Saturn composition planets in the mass range 50 – 100 $M_\oplus$.

**Table 3: Rock Terrestrial and Rock-Ice Terrestrial Planets ($f_{H-He}$ < 0.01% by mass)**

| Planet | Period | Mass | Radius | Flux / Temp | Reference |
|---|---|---|---|---|---|
| TRAPPIST-1 h | 18.77 | 0.326 +/-0.02 | 0.755 +/-0.14 | 0.144 / - | Agol et al. 2021 |
| TRAPPIST-1 d | 4.05 | 0.388 +/-0.012 | 0.788 +.011/-.010 | 1.115 / - | Agol et al. 2021 |
| TRAPPIST-1 e | 6.10 | 0.692 +/-0.022 | 0.920 +.013/-.012 | 0.646 / - | Agol et al. 2021 |
| TRAPPIST-1 f | 9.21 | 1.039 +/-0.031 | 1.045 +.013/-.012 | 0.373 / - | Agol et al. 2021 |
| TRAPPIST-1 c | 2.42 | 1.308 +/-0.056 | 1.097 +.014/-.012 | 2.1214 / - | Agol et al. 2021 |
| TRAPPIST-1 g | 12.35 | 1.321 +/-0.038 | 1.129 +.015/-.013 | 0.252 / - | Agol et al. 2021 |
| TRAPPIST-1 b | 1.51 | 1.374 +/-0.069 | 1.116 +.014/-.012 | 4.153 / - | Agol et al. 2021 |
| LTT 1445 A c | 3.12 | 1.54 +.20/-.19 | 1.147 +.055/-.054 | 10.9 / 508 | Winters et al. 2022 |
| TOI-270 b | 3.36 | 1.58 +/-0.26 | 1.206 +/-0.039 | - / 581 | Van Eylen et al. 2021 |
| GJ 1132 b | 1.63 | 1.66 +/-0.23 | 1.13 +/-0.056 | - / 529 | Bonfils et al. 2018 |
| GJ 3929 b | 2.62 | 1.75 +.44/-.45 | 1.09 +/-0.04 | 17.2 / 568 | Beard et al. 2022 |
| GJ 357 b | 3.93 | 1.84 +/- 0.31 | 1.217 +.084/-.083 | 12.6 / 525 | Luque et al. 2019 |
| L98-59 c | 3.69 | 2.22 +.26/-.25 | 1.385 +.095/-.075 | 12.8 / 553 | Demangeon et al. 2021 |
| LHS 1478 b | 1.95 | 2.33 +/-0.20 | 1.242 +.051/-.049 | - / 595 | Soto et al. 2021 |
| LTT 1445 A b | 5.36 | 2.87 +.26/-.25 | 1.305 +.066/-.061 | 5.4 / 424 | Winters et al. 2022 |
| LHS 1140 b | 24.74 | 6.38 +.46/-.44 | 1.635 +/-0.046 | - / 379 | Lillo-Box et al. 2020 |

**Table 4: Gas-rich, Secondary Envelope, or Supercritical Hydrosphere Terrestrial Planets**

| Planet | Period | Mass | Radius | Flux / Temp | Reference |
|---|---|---|---|---|---|
| L98-59 d | 7.45 | 1.94 +/- 0.28 | 1.521 +.119/-.098 | 5.01 / 416 | Demangeon et al. 2021 |
| Kepler 54 c | 12.07 | 2.10 +.21/-.20 | 1.688 +.054/-.055 | 7.165* / - | Leleu et al. 2023 |
| Kepler 138 c | 13.78 | 2.3 +0.6/-0.5 | 1.51 +/-0.04 | 6.8 / 410 | Piaulet et al. 2023 |
| HD 260655 c | 5.71 | 3.09 +/-0.48 | 1.533 +.051/-.046 | 16.1 / 557 | Luque et al. 2022 |
| Kepler 54 b | 8.01 | 3.09 +.30/-.31 | 1.856 +/-0.057 | 12.357*/ - | Leleu et al. 2023 |
| HD 23472 c | 29.80 | 3.41 +.88/-.81 | 1.87 +0.12/-0.11 | 7.96 / 467 | Barros et al. 2022 |
| TOI-776 b | 8.25 | 4.0 +/-0.9 | 1.85 +/-0.13 | 11.5 / 514 | Luque et al. 2021 |
| G9-40 b | 5.75 | 4.00 +/-0.63 | 1.900 +/-0.065 | 6.27 / 441 | Luque et al. 2022 |
| TOI-270 d | 11.38 | 4.78 +/-0.43 | 2.133 +/-0.058 | - / 387 | Van Eylen et al. 2021 |
| TOI-1452 b | 11.06 | 4.82 +/-1.30 | 1.672 +/-0.071 | 1.8 / 326 | Cadieux et al. 2022 |
| K2-146 b | 2.64 | 5.77 +/-0.18 | 2.05 +/-0.06 | 20.7 / 534 | Hamann et al. 2019 |
| TOI-1468 c | 15.53 | 6.64 +/-0.68 | 2.06 +/-0.044 | 2.15 / 338 | Chaturvedi et al. 2022 |
| K2-146 c | 4.00 | 7.49 +/-0.24 | 2.19 +/-0.07 | - / <534 | Hamann et al. 2019 |
| HD 23472 b | 17.67 | 8.32 +.78/-.79 | 2.00 +0.11/-0.10 | 16.0 / 543 | Barros et al. 2022 |
| Kepler 538 b | 81.74 | 10.6 +2.5/-2.4 | 2.215 +.040/-.034 | 2.99 / 380 | Mayo et al. 2019 |
| K2-263 b | 50.82 | 14.8 +/-3.1 | 2.41 +/-0.12 | - / 470 | Mortier et al. 2018 |
| GJ143 b | 35.61 | 22.7+2.7/-1.9 | 2.61 +0.17/-0.16 | - / 422 | Dragomir et al. 2019 |

**Table 5: Rock-Ice Giants (Neptunes - $f_{H-He}$ 1.0 – 49.9 % by mass)**

| Planet | Period | Mass | Radius | Flux / Temp | Reference |
|---|---|---|---|---|---|
| **HD 191939 d** | 38.35 | 2.8 +/-0.6 | 2.995 +/-0.07 | 14.3 / 540 | Orell-Miguel et al. 2023 |
| **Kepler 289 d** | 66.06 | 4.0 +/-0.9 | 2.68 +/-0.17 | | Schmitt et al. 2014 |
| **Kepler 26 b** | 12.28 | 4.85 +.44/-.42 | 3.22 +/-0.15 | 10.45* / - | Leleu et al. 2023 |
| **Kepler 51 d** | 130.18 | 5.70 +/-1.12 | 9.46 +/-0.16 | 2.466* / - | Libby-Roberts et al. 2020 |
| **TOI-270 c** | 5.66 | 6.15 +/-0.37 | 2.355 +/-0.064 | - / 488 | Van Eylen et al. 2021 |
| **Kepler 305 d** | 16.74 | 6.20 +1.76/-1.34 | 2.76 +/-0.12 | 25.52* / - | Leleu et al. 2023 |
| **LTT 3780 c** | 12.25 | 6.29 +.63/-.61 | 2.42 +/-0.10 | 2.88 / 397 | Nowak et al. 2020 |
| **Kepler 79 e** | 81.06 | 6.3 +/-1.0 | 3.414 +/-0.129 | 18.78* / - | Yofee et al. 2021 |
| **Kepler 87 c** | 191.23 | 6.4 +/-0.8 | 6.14 +/-0.29 | - / 403 | Ofir et al. 2014 |
| **Kepler 10 c** | 45.29 | 7.37[a]+1.32/-1.19 | 2.32 +.09/-.08 | 18.300* / - | Weiss et al. 2016 |
| **Kepler 26 c** | 17.25 | 7.48 +.49/-.48 | 3.11 +/-0.14 | 6.63* / - | Leleu et al. 2023 |
| **TOI-178 f** | 15.23 | 7.72 +1.67/-1.62 | 2.287 +/-0.11 | - / 521 | Leleu et al. 2021 |
| **HD 191939 c** | 28.58 | 8.0 +/-0.1 | 3.195 +/-0.075 | 21.0 / 600 | Orell-Miguel et al. 2023 |
| **GJ 1214 b** | 1.58 | 8.17 +/-0.43 | 2.742 +/-.05 | 21.0 / 596 | Cloutier et al. 2021 |
| **Kepler 49 c** | 10.91 | 8.38 +.92/-.89 | 2.444 +.083/-.082 | 12.34* / - | Leleu et al. 2023 |
| **K2-18 b** | 32.94 | 8.63+/-1.35 | 2.610 +/-0.087 | 1.005 / 255 | Benneke et al. 2019 |
| **TOI-269 b** | 3.70 | 8.80 +/-1.40 | 2.77 +/-0.12 | 19.0 / 531 | Cointepas et al. 2021 |
| **HD 136352 d** | 107.25 | 8.82 +.93/-.92 | 2.562 +.088/-.079 | 5.74 / 431 | Delraz et al. 2021 |
| **HD 73583 c** | 18.88 | 9.70 +1.80/-1.70 | 2.39 +.10/-.09 | 10.2 / 498 | Barragán et al. 2022 |
| **Kepler 49 b** | 7.20 | 9.77 +.94/-.95 | 2.579 +.087/-.086 | 21.48* / - | Leleu et al. 2023 |
| **HD 3167 c** | 29.85 | 10.67 +.85/-.81 | 2.923 +.098/-.109 | - / 565 | Bourrier et al. 2022 |
| **TOI-1759 b** | 18.85 | 10.8 +/-0.15 | 3.14 +/-0.10 | 6.39 / 443 | Espinoza et al. 2022 |
| **Kepler 82 b** | 26.44 | 12.15 +.96/-.87 | 4.07 +.24/-.10 | 19.558* / - | Freudenthal et al. 2019 |
| **GJ 3470 b** | 3.34 | 12.58 +/-1.3 | 3.88 +/-0.32 | - / 615 | Kosiarek et al. 2019 |
| **Kepler 82 c** | 51.54 | 13.9 +1.3/-1.2 | 5.34 +.33/-.13 | 8.035* / - | Freudenthal et al. 2019 |
| **TOI-1246 e** | 37.92 | 14.8 +/-2.3 | 3.78 +/-0.16 | 11 / 462 | Turtelboom et al. 2022 |
| **TOI-1231 b** | 24.25 | 15.4 +/-3.3 | 3.65 +.16/-.15 | 1.93 / 330 | Burt et al. 2021 |
| **K2-24 c** | 42.34 | 15.4 +1.9/-1.8 | 7.5 +/-0.3 | 24 / 606 | Petigura et al. 2018 |
| **TOI-561 e** | 77.23 | 16.0 +/-2.3 | 2.67 +/-0.11 | - / <600K | Lacadelli et al. 2020 |
| **TOI-216 b** | 17.10 | 17.7 +0.7/-0.6 | 7.28 +0.14/-0.13 | 25.9** / - | McKee & Montet 2022 |
| **K2-25 b** | 3.48 | 24.5 +5.7/-5.2 | 3.44 +/-0.12 | 9.91 / 494 | Steffansson et al. 2020 |
| **TOI-3884 b** | 4.54 | 32.59 +7.31/-7.38 | 6.43 +/-0.20 | 6.29 / 441 | Libby-Roberts et al. 2023 |
| **HD 95338 b** | 55.09 | 42.44 +2.22/-2.08 | 3.89 +.19/-.20 | 7.42 / 385 | Díaz et al. 2020 |

*Planets with stellar flux from Berger et al. 2018
**Stellar flux from Kipping et al. 2019
Notes: (a) Kepler 10 c mass from Rajpaul et al. 2017 and radius from Weiss et al. 2016.

**Table 6: Gas Giants (Saturns - $f_{H-He}$ 50.1 – 80.0 % by mass)**

| Planet | Period | Mass | Radius | Flux / Temp | Reference |
|---|---|---|---|---|---|
| **Kepler 177 c** | 49.41 | 14.7 +2.7/-2.5 | 8.73 +.36/-.34 | 25.4 / - | Vissapragada et al. 2020 |
| **Kepler 30 d** | 143.34 | 23.1 +/-2.7 | 8.8 +/-0.5 | 1.962* / - | Sanchis-Ojeda et al. 2012 |
| **Kepler 9 c** | 38.99 | 29.9 +1.1/-1.3 | 8.08 +.54/-.41 | 18.903* / - | Borsata et al. 2019 |
| **Kepler 35 b** | 131.46 | 40.36 +/-6.356 | 8.15 +/-0.157 | | Welsh et al. 2012 |
| **TOI-3984 A b** | 4.35 | 44.0 +8.7/-8.0 | 7.9 +/-0.24 | - / 563 | Cañas et al. 2023 |
| **Kepler 34 b** | 288.82 | 69.92 +3.5/-3.2 | 8.564 +.135/-.157 | | Welsh et al. 2012 |
| **KOI-1780.01** | 134.46 | 71.0 +11.2/-9.2 | 8.86 +.25/-.24 | 5.70 / - | Vissapragada et al. 2020 |
| **Kepler 16 b** | 228.78 | 105.83 +/-5.08 | 8.449 +/-0.029 | | Doyle et al. 2011 |
| **NGTS-11 b** | 35.46 | 109 +29/-23 | 9.16 +.31/-.36 | - / 435 | Gill et al. 2020 |
| **K2-139 b** | 28.83 | 123 +26/-24 | 9.06 +.38/-.37 | - / 565 | Barragán et al. 2018 |
| **TOI-530 b** | 6.39 | 127 +29/-32 | 9.3 +/-0.7 | - / 565 | Gan et al. 2022 |
| **TOI-216 c** | 34.55 | 168 +/-6 | 10.08 +/-.09 | 10.10 / 497 | McKee & Montet 2022 |

*Planets with stellar flux from Berger et al. 2018

**Stellar flux from Kipping et al. 2019

**Table 7: Gas Giants (Jupiters - $f_{H-He}$ 80.1 – 100.0 % by mass)**

| Planet | Period | Mass | Radius | Flux / Temp | Reference |
|---|---|---|---|---|---|
| **Kepler 289 c** | 125.85 | 132 +/-17 | 11.59 +/-0.19 | 4.882* / - | Schmitt et al. 2014 |
| **TOI-3235 b** | 2.59 | 211.3 +/-7.95 | 11.16 +/-0.48 | - / 604 | Hobson et al. 2023 |
| **TOI-1899 b** | 29.09 | 212.9 +/-12.71 | 10.86 +/-0.33 | - / 378 | Lin et al. 2023 |
| **CoRoT-9 b** | 95.27 | 267 +/-16 | 11.95 +.84/-.71 | - / 420 | Bonomo et al. 2017 |
| **Kepler 167 e** | 1071.23 | 321 +54/-51 | 10.16 +/-0.42 | - / 134 | Chachan et al. 2022 |
| **Kepler 87 b** | 114.74 | 324.2 +/-8.8 | 13.49 +/-0.55 | - / 478 | Ofir et al. 2014 |
| **TOI-5542 b** | 75.12 | 420 +/-32 | 11.31 +.40/-.39 | 9.60 / 441 | Grieves et al. 2022 |
| **KOI-3680 b** | 141.24 | 613 +60/-67 | 11.1 +.7/-.8 | - / 347 | Hébrard et al. 2019 |
| **Kepler 30 c** | 60.32 | 640 +/-50 | 12.3 +/-0.4 | 6.217* / - | Sanchis-Ojeda et al. 2012 |
| **TOI-4562 b** | 225.12 | 732 +152/-149 | 12.53 +/-0.15 | - / 318 | Heitzmann et al. 2023 |
| **CoRoT-10 b** | 13.24 | 874.00 +/-50.85 | 10.87 +/-0.70 | - / 600 | Bonomo et al. 2010 |
| **TOI-2180 b** | 260.79 | 875.6 +27.7/-25.7 | 11.32 +.25/-.24 | 2.71 / 348 | Dalba et al. 2022 |
| **HD 80606 b** | 111.44 | 1308 +/-49 | 11.24 +/-0.30 | - / 405 | Southworth et al. 2011 |
| **Kepler 1704 b** | 988.88 | 1319 +/-92 | 11.94 +.48/-.46 | 0.342 / 254 | Dalba et al. 2021 |
| **Kepler 1514 b** | 217.83 | 1678 +/-70 | 12.42 +/-0.26 | 3.23 / 388 | Dalba et al. 2021 |

*Planets with stellar flux from Berger et al. 2018

**Table 8: Radius Trends for < 600 K sample**

| Composition | Radius Range ($R_\oplus$) | Radius Range Exceptions |
|---|---|---|
| **Terrestrial** | < 1.5 | LHS 1140 b ($R_\oplus$ = 1.635) |
| **Gas-rich Terrestrial** | 1.50 – 2.25 | K2-263 b ($R_\oplus$ = 2.41); GJ 143 b ($R_\oplus$ = 2.61) |
| **Gas-poor Neptune** | 2.25 – 3.00 | |
| **Neptune** | 3.0 – 4.5 | |
| **Gas-rich Neptune** | 4.5 – 7.5 | Kepler 51 d ($R_\oplus$ = 9.46) |
| **Saturn** | 7.5 – 10.0 | |
| **Jupiter** | 10.0 – 13.5 | |

~4. Catalog Sample Considered in the Context of Other Exoplanet Results

In this section, the characteristics of the current catalog sample of exoplanets with Teq ≤ 600 K are considered in relation to the results of other exoplanet studies. This discussion is meant to illustrate connections to a few relevant topics of exoplanet study rather than be an exhaustive review. It should also be noted that the total sample size meeting the selection criteria of this catalog is 93 planets. When considering the smaller numbers for sub-samples of different composition classes it is possible that some of the conclusions are effected by small number statistics.

For example, TOI-216 c ($M_\oplus$ = 168) was recently analyzed by McKee & Montet (2022) and has the largest mass of the Saturn composition planets (Table 6). TOI-216 c is currently the single example in this catalog of a Saturn composition planet that has more mass than a Jupiter composition planet (Kepler 289 c - $M_\oplus$ = 132). A similar example is found with TOI-3884 b, a 6.0 $R_\oplus$ planet that was found to have a mass of 16.5 $M_\oplus$ (Almenara et al. 2022) indicating the planet is a gas-rich Neptune. However, in a very recent analysis Libby-Roberts et al. (2023) find that the mass of TOI-3884 b is 32.59 $M_\oplus$ with a radius of 6.43 $R_\oplus$ and therefore the planet is a gas-rich super-Neptune.

As illustrated by TOI-3884 b and TOI-216 c, it is possible that continuing exoplanet discoveries and improved mass and radius measurements for currently known exoplanets will alter some of the population statistics and mass ranges described in this first version of the catalog. With this caution noted, the sections that follow describe connections between the sample in this catalog and the ongoing study of exoplanets in general.

~4.1 Terrestrial planets

~4.1.1 Terrestrial Planet Core Mass Fractions

Zeng et al. (2016) provided mass-radius models for characterizing Terrestrial planets from 100% Fe to 100% $H_2O$ composition. Applying these models, the Terrestrial planets in the Teq ≤ 600 K catalog (Table 3) include: 1 planet with $f_{H2O}$ 0 – 10%, 5 planets with $f_{Fe}$ 0 – 20%, 8 planets with $f_{Fe}$ 20 – 40%, and 2 planets with $f_{Fe}$ >40%. Spaargaren et al. (2022) found that Terrestrial planets in the Solar neighborhood have Fe core mass fractions from 18 – 35% by mass. At least half the Terrestrial planets in this catalog therefore have core mass fractions consistent with the findings of Spaargaren et al. (2022).

Five Terrestrial planets in the catalog are consistent with 5 – 15% by mass Fe cores. Due to model degeneracy, these planets could have an alternative composition with a higher Fe core fraction, in the range determined by Spaargaren et al. (2022), combined with a low percent by mass $H_2O$ layer. For example, TRAPPIST-1 f could have a 15% Fe core from the models of Zeng et al. (2016), but the mass and radius values are also consistent with an alternative composition that has Fe core and $H_2O$ mass fractions of ~35% and ~10% respectively. This composition is reasonable for TRAPPIST-1 f as the stellar flux is only 0.373 $S_\oplus$ (Teq ≈ 220K), the radius is below that expected for a planet with a 0.01% supercritical hydrosphere from the models of Vivien et al. (2022), and therefore an inflated supercritical hydrosphere is unlikely.

~4.1.2 Mercury Composition Planets

It has remained challenging to determine the formation channels for Mercury (Fe core mass fraction ~70%) and the growing population of super-Mercuries (Fe core fraction >50%) identified in the exoplanet population. One explanation is the "Giant Impact" (GI) hypothesis in which Mercury composition planets result from mantle stripping due to GI in the later stages of planetary assembly (e.g. Spalding & Adams 2020; Reinhardt et al. 2022). Another explanation for super-Mercuries is that a high Fe core fraction can result from an enhanced Fe composition in the proto-planetary disk as indicated by a correlation between super-Mercuries and the composition of their stellar hosts (Adibekyan et al. 2021). Johansen & Dorn (2022) explored the conditions for the nucleation and growth of iron-rich pebbles which could provide a corresponding mechanism for producing Mercury composition planets around iron-rich stars in the absence of GI. See discussion in Adibekyan et al. (2021) and Reinhardt et al. (2022) for review of the strengths and challenges faced by both the GI and stellar composition explanations for super-Mercuries.

Reinhardt et al. (2022) provided a list of 9 iron-rich exoplanets in their analysis of the role of GI on super-Mercury formation which includes all 5 super-Mercuries previously identified by Adibekyan et al. (2021). Two additional

Mercuries with masses < 0.8 $M_⊕$ were recently reported by Barros et al. (2022) bringing the total to eleven. All of these planets have Teq ≥ 720 K and 7 have Teq >1150 K. Among the Terrestrial planets listed in Table 3 of this catalog, the planet GJ 3929 b (Beard et al. 2022) has a mass and radius consistent with a Mercury composition of ~60% Fe core from the models of Zeng et al. (2016). GJ 3929 b has the highest Fe core mass fraction among the Teq ≤ 600K sample in this catalog, and has the lowest equilibrium temperature among all potential Mercury composition exoplanets identified to date.

It should be noted that in a different recent analysis Kemmer et al. (2022) found a smaller mass and slightly larger radius for GJ 3929 b than the values found by Beard et al. (2022), but with larger uncertainty in the planetary mass than the standard for inclusion in this catalog. The Kemmer et al. (2022) mass and radius values match a composition that is ~5% $H_2O$ on a pure silicate rock core (Zeng et al. 2016) which provides a cautionary example that illustrates the large range of possible compositions any given exoplanet could have based upon different analyses reported in the literature. With that caveat, GJ 3929 b orbits the lowest mass star among potential Mercuries and is the only Terrestrial planet in the Teq ≤600K sample that currently has reported mass and radius values consistent with a Mercury composition.

### ~4.1.3 Terrestrial Planet Mass Distribution

The sample of Terrestrial planets in this catalog provides an interesting contrast in mass distribution to Terrestrial planet samples that include much higher stellar flux values (Dai et al. 2019; Otegi et al. 2020; Reinhardt et al. 2022). With the exception of LHS 1140 b ($M_⊕$ = 6.38), the remaining 15 Terrestrial planets in Table 3 all have a mass < 2.9 $M_⊕$. It is important to note that 3 $M_⊕$ is less than the minimum critical core mass, ~4 $M_⊕$, studies of the core accretion model find is required to initiate accumulation of a significant H-He envelope (e.g. Ida & Lin 2004).

The mass distribution for Terrestrial composition planets in this Teq ≤ 600K catalog differs from the Mercury composition planets included in Reinhardt et al. (2022) which have masses as large as 9.39 $M_⊕$ with 7 of the 9 planets having a mass > 3.5 $M_⊕$. Similarly, Dai et al. (2019) analyzed a sample of twelve ultra-short period (p < 1 day) Terrestrial planets with 10 of the 12 planets having a mass > 3.5 $M_⊕$. Dai et al. (2019) note that the high stellar flux values for their sample (>650 $S_⊕$) would strip any H-He envelope and suggest that these hot Earths are the exposed rocky cores of sub-Neptunes. In this context, it is possible that the Terrestrial planet sample in the catalog presented here may represent a different planet population with different formation pathways, and certainly different evolutionary histories, than the highly irradiated planets analyzed in other studies (e.g. Dai et al. 2019; Reinhardt et al. 2022). The Terrestrial planets in this catalog may not be the exposed cores of sub-Neptunes and should be compared against the predictions of models that form Terrestrial planets without substantial H-He envelopes.

LHS 1140 b has the longest period (~25 days), largest mass (6.38 $M_⊕$), largest radius (1.635 $R_⊕$), and is the only Terrestrial composition planet in this sample with a mass exceeding the core accretion critical core mass for accumulating a substantial H-He envelope. The large gap between the mass of LHS 1140 b and the next largest planet (LTT 1445 A b) in the Terrestrial sample is noteworthy since LHS 1140 b is over twice the mass of LTT 1445 A b (Figure 1). There is no similar mass jump in the Mercury composition population (Reinhardt et al. 2022) or in the USP hot Earth sample (Dai et al. 2019). The mass and radius values for LHS 1140 b are consistent with a 40% Fe core (Zeng et al. 2016). The equilibrium temperature of LHS 1140 b is 379 K and therefore it lacks the high stellar flux that would suggest a sub-Neptune core that has been exposed by photoevaporation based upon the models of Lopez & Fortney (2013). Instead, the lack of an H-He envelope for LHS 1140 b might be explained by core powered mass loss (e.g. Ginzburg et al. 2018), giant impacts (e.g. Reinhardt et al. 2022), formation in a gas poor environment (e.g. Lee & Chiang 2016), or a combination of these mechanisms. If massive Terrestrial planets frequently form with a negligible H-He envelope then similar low Teq examples to LHS 1140 b should eventually be discovered with masses between 3 and 6 $M_⊕$.

### ~4.1.4 Gas-rich Terrestrial and Supercritical Hydrosphere Composition Planets

Recent models have explored the mass-radius relationships for water rich Terrestrial planets irradiated sufficiently to develop a supercritical hydrosphere (Mousis et al. 2020; Aguichine et al. 2021; Turbet et al. 2021; Vivien et al. 2022). These studies demonstrate that many planets with mass and radius consistent with the presence of an extended

H-He envelope could instead be H-He envelope free but with a supercritical hydrosphere and radius inflated with a steam envelope. The possibility that some Terrestrial planets possess a supercritical hydrosphere presents a significant degeneracy in determining planetary composition because the planets in question could also possess envelopes with $f_{H-He}$ 0.1 – 1.0% by mass.

All of the Terrestrial planets in Table 3 of this catalog have radii too small to possess a substantial primordial H-He envelope since the $f_{H-He}$ would be <0.01% by mass from the models of Lopez & Fortney (2014). The planets in Table 3 also have radii smaller than required to have a 0.01% by mass hydrosphere following the models of Vivien et al. (2022) assuming a core mass fraction (CMF) of zero percent by mass. In a more realistic model, with a CMF ≈ 30 %, the predicted radii for supercritical hydrosphere Terrestrial planets are expected to be smaller by .06 - .08 $R_\oplus$ based upon the radii of planets from 1 – 3 $M_\oplus$ from the models of Zeng et al. (2016). With a CMF ≈ 30%, the Terrestrial planet L98-59 c could potentially possess a supercritical hydrosphere of ~1% by mass from the model of Vivien et al. (2022). The remaining Terrestrial planets in Table 3 have radii smaller than predicted for a 0.01% by mass supercritical hydrosphere even if the CMF ≈ 30%.

The planets in Table 3 have radii too small to have a significant primordial H-He envelope or a supercritical hydrosphere but these planets could have a secondary atmosphere from mantle outgassing (Herbort et al. 2020; Grenfell et al. 2020). This secondary atmosphere could include primordial $H_2$ that was absorbed by the mantle and later outgassed (e.g. Grenfell et al. 2020). For example, GJ 1132 b (Table 3) should have lost any primordial H-He atmosphere, but was initially found from Hubble Space Telescope observations to have spectral signatures consistent with a secondary outgassed $H_2$, $CH_4$, HCN and aerosol atmosphere (Swain et al. 2021). However, a follow-up study by Mugnai et al. (2021) could not confirm the spectral signatures identified for GJ 1132 b in Swain et al. (2021). While these results are conflicting, they illustrate the type of observations that are necessary to characterize the atmospheres of Terrestrial planets such as those in Table 3 that are too small to have a substantial primordial H-He envelope or a supercritical hydrosphere.

The potential gas-rich Terrestrial planets in Table 4 have mass-radius values consistent with $f_{H-He}$ 0.1 – 0.9% by mass using the Lopez & Fortney (2014) models, or supercritical hydrospheres with 5 – 20% water using the models of Aguichine et al. (2021) and Vivien et al. (2022). In order to break the degeneracy in these models atmospheric characterization will be necessary. For example, Mikal-Evans et al. (2023) recently used transmission spectroscopy to study TOI-270 d (Table 4) and found that the planet has a hydrogen-rich envelope rather than a steam envelope. However, TOI-270 d could also have a "Hycean" planet composition with a hydrogen envelope over a water ocean (Mikal-Evans et al. 2023). Possible characterizations for TOI-270 d therefore include: "*gas-rich super-Earth*", "*gas-rich super-Europa*", or "*gas-rich super-Ganymede*" composition.

Most of the planets in Table 4 have sub-Neptune mass values from 1.9 – 8.5 $M_\oplus$ and sub-Neptune radius values from 1.50 – 2.20 $R_\oplus$. The lowest mass gas-rich Terrestrial planet identified for this catalog is L98-59 d ($M_\oplus$ = 1.94) and is consistent with a $f_{H-He}$ 0.1 – 0.2% from Lopez & Fortney (2014). If L98-59 d instead has a supercritical hydrosphere then the $H_2O$ mass fraction is ~4 – 5 % with CMF = 0 from the model of Vivien et al. (2022) or ~5 – 6 % for CMF ≈ 30 when extrapolating the model of Vivien et al. (2022) based upon the radii predicted for different CMF from Zeng et al. (2016). From Hubble spectroscopy Zhou et al. (2023) conclude that L98-59d cannot have a clear hydrogen dominated atmosphere, but could have a cloudy hydrogen dominated atmosphere, although their model favors a cloudy secondary atmosphere.

K2-263 b and GJ 143 b have radii of 2.41 and 2.61 $R_\oplus$ respectively and are the only two gas-rich Terrestrial planets in Table 4 that overlap the radius range for gas-poor Neptune composition planets (Figure 1). These planets have Neptune rather than sub-Neptune mass values and $f_{H-He}$ ~0.8 – 0.9% from the models of Lopez & Fortney (2014). Given the low equilibrium temperatures (< 475 K) and large masses for these planets it is unlikely that photoevaporation can explain the low $f_{H-He}$ (Lopez & Fortney 2013). But based upon the core accretion scenario (Pollack et al. 1996; Ida & Lin 2004) an explanation is needed for how K2-263 b and GJ 143 b acquired such large masses without also accumulating a substantial H-He envelope.

Roy et al. (2022) recently identified the dense sub-Neptune TOI-824 b as a possible exposed Neptune mantle based upon eclipse depths, which indicate a high metallicity atmosphere. TOI-824 b has a similar mass (18.47 $M_\oplus$) to K2-263 b and GJ 143 b but with larger radius (2.93 $R_\oplus$), and much higher equilibrium temperature (1253 K – Burt et al. 2020). TOI-824 b is consistent with having a 1 – 2 % by mass H-He envelope from Lopez & Fortney (2014) and therefore provides an interesting comparison case.  Roy et al. (2022) find that if TOI-824 b started with 1 – 10% by mass Hydrogen layers, photoevaporation would result in minimal atmospheric loss and most of the H-He would be retained.  Given that K2-263 b and GJ 143 b both have an equilibrium temperature < 500 K, it should be even more difficult for photoevaporation to be responsible for the low percentage H-He for these planets.

Roy et al. (2022) suggest that atmospheric mass loss from giant impacts could explain the low H-He mass fraction for TOI-824 b.  The thermal energy released from giant impacts could drive partial, or complete, H-He mass loss and is one possible explanation for the low H-He mass fractions for the massive gas-rich super-Earths K2-263 b, GJ 143 b, and Kepler 538 b (10.6 $M_\oplus$ - Mayo et al. 2019) in this catalog.  An alternative explanation for Neptune mass gas-rich Terrestrial composition planets is formation in a gas poor environment (Lee & Chiang 2016).

### ~4.2 Rock-Ice Giants (Neptunes)

Approximately 1/3 of the Teq ≤ 600 K sample in this catalog are classified as Rock-Ice Giants or "Neptunes". The population characteristics for this sample are summarized in Table 9.  The distinction between the gas composition ranges and mass ranges is important.  While most of the gas-poor Neptunes are sub-Neptune mass (14/16), there is a relevant fraction of the sub-Neptune mass planet population (< 10 $M_\oplus$) that has a Neptune or gas-rich Neptune composition (6/20 or 30%).

The mass distribution of this Rock-Ice Giant sample contains several general characteristics and examples of interest. First, including Uranus and Neptune, only 3/35 low temperature Neptunes have a mass exceeding 18 $M_\oplus$. Neptune and Uranus are respectively the 5$^{th}$ and 10$^{th}$ most massive Rock-Ice Giants in the sample.  This limited sample of Rock-Ice Giants with mass exceeding 18 $M_\oplus$ is not found in the sample of Otegi et al. (2020), which includes seven Rock-Ice Giants with mass from 19 – 30 $M_\oplus$ and an additional seven Rock-Ice Giants with mass from 30 – 60 $M_\oplus$. Most of the planets in the Otegi et al. (2020) sample have significantly higher stellar flux values than the planets in this sample.

The population of Rock-Ice Giants with Teq <600 K nearly truncates at 18 $M_\oplus$ (Figure 1).  This may result from formation mechanisms such as core accretion which predicts a critical mass to be reached by ~20 $M_\oplus$ resulting in runaway gas accretion when the fraction of gas is approximately equal to the fraction of rock and ice (Pollack et al. 1996; Ida & Lin 2004; see recent review of planet formation mechanisms by Raymond et al. 2020).  The small number of low Teq Neptunes exceeding 20 $M_\oplus$ is consistent with this prediction of the core accretion scenario.  It should be noted that 20 $M_\oplus$ is also the pebble isolation mass for cores grown by pebble accretion (Lambrechts et al. 2014).  The higher number of Rock-Ice Giants with mass > 20 $M_\oplus$ found in the shorter orbital period and higher stellar flux sample of Otegi et al. (2020) could be the result of different formation pathways and environments or post-formation evolutionary effects resulting from inflationary mechanisms and higher stellar flux in general.

**Table 9: Characteristics of Rock-Ice Giant Sample**

| Characterization | Criteria | Number* | fraction |
|---|---|---|---|
| **Sub-Neptune mass** | 2.5 – 10.0 $M_\oplus$ | 20/35 | .571 |
| **Neptune mass** | 10.0 – 30.0 $M_\oplus$ | 13/35 | .371 |
| **Super-Neptune** | >30.0 $M_\oplus$ | 2/35 | .057 |
| | | | |
| **Gas-Poor Neptune** | $f_{H-He}$ 1 – 4% , $R_\oplus$ 2.25 – 3.0 | 16/35 | .457 |
| **Neptune** | $f_{H-He}$ 4 – 20 , $R_\oplus$ 3.0 – 4.5 | 13/35 | .371 |
| **Gas-Rich Neptune** | $f_{H-He}$ 20 – 49.9 , $R_\oplus$ >4.5 | 6/35 | .171 |

*Total includes Table 5 sample, Uranus, and Neptune

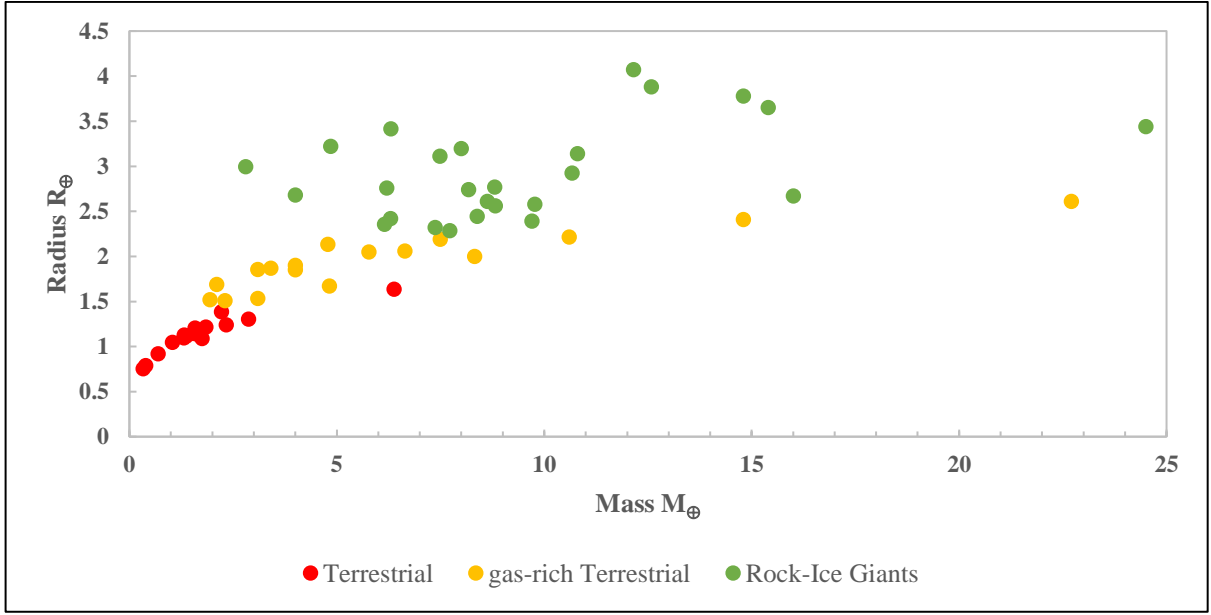

**Figure 1:** Mass-Radius diagram for planets in the catalog with mass < 25 M$_⊕$ and radius < 4.5 R$_⊕$.

Second, HD 191939 d is the least massive Rock-Ice Giant with a mass of 2.8 M$_⊕$ and a radius of 2.995 R$_⊕$ (Orell-Miguell et al. 2023). From Lopez & Fortney (2014) HD 191939 d has a 3 – 4 % by mass H-He envelope and can be characterized as a "*gas-poor sub-Neptune*". It is important to note that the lower mass limit for Rock-Ice giants in this sample (2.8 M$_⊕$) corresponds with the upper mass limit for all but one of the Terrestrial planets (section 4.1.1). This may be seen as further support for the core accretion critical core mass prediction (e.g. Ida & Lin 2004) that cores > ~4.0 M$_⊕$ should begin to accrete a substantial H-He envelope since the rest of the planets in the Rock-Ice Giant sample have a mass of at least 4.0 M$_⊕$.

Third, it is interesting to note that while mass cannot be used to predict the H-He fraction for this T ≤ 600K sample of Neptunes, the planetary radius can be used as a proxy for composition as described in section 3.3. The strong correlation between radius and composition in this sample, as compared with higher Teq samples, most likely results from the reduced influence of evolutionary and inflationary effects from stellar irradiation and tidal inflation compared to other samples. For example, Armstrong et al. (2020) found that TOI-849 b (M$_⊕$ = 39.09, R$_⊕$ = 3.444) has $f_{H-He}$ < 4% and therefore this planet is a gas-poor Neptune. TOI-849 b does not follow the "radius as a proxy for composition" trends found for the sample in this paper (Table 8). Planets in this catalog with similar radius to TOI-849 b all have a Neptune composition, rather than the gas-poor Neptune composition of TOI-849 b. The high stellar flux value of this planet may explain this difference as TOI-849 b has a 0.765 day period orbiting a star with an effective temperature of ~5300 K, and therefore has an equilibrium temperature > 1700 K (Armstrong et al. 2020). In contrast to TOI-849 b, the similar mass and radius planet from this catalog, HD 95338 b (M$_⊕$ = 42.44, R$_⊕$ = 3.89), is estimated by Diaz et al. (2020) to have a $f_{H-He}$ of ~10% and therefore is a Neptune composition Rock-Ice Giant with its radius predicting the composition independent of the mass of the planet.

### ~4.3 Gas Giants (Saturns and Jupiters)

The $f_{H-He}$ is used to separate Gas Giants into Saturns ($f_{H-He}$ 50.0 – 80.0 % by mass) and Jupiters ($f_{H-He}$ 80.1 – 100.0 % by mass). The most intriguing planets in this sample are the three Saturn composition planets with radii from 8.08 – 8.8 R$_⊕$ that have Neptune masses from 14.7 – 29.9 M$_⊕$. Based upon the models of Fortney et al. (2007) these low mass Saturns should have $f_{H-He}$ in the range 55 – 65% by mass. In the core accretion model, runaway gas accretion occurs when the H-He gas fraction ≈ core fraction (Pollack et al. 1996; Ida & Lin 2004; review by Raymond et al. 2020). Kepler 177c has a mass of 14.7 M⊕ which might suggest a minimum core mass for runaway accretion of ~7.5 M⊕. This is consistent with the findings of Dai et al. (2019) who identified an upper mass of 8 M⊕ for USP Earths which were interpreted as the exposed cores of sub-Neptunes. Dai et al. (2019) suggested this mass limit as the threshold mass limit for cores of planets that experience runaway gas accretion in agreement with Lee & Chiang (2016) and Piso et al. (2015).

As discussed in section 3.4, the second intriguing aspect of the Gas Giant population with Teq ≤ 600 K is the lack of a notable mass overlap between Saturn and Jupiter composition Gas Giants with all but one Saturn having a mass < 130 M⊕ and every Jupiter having a mass > 130 M⊕. This is in contrast to the Otegi et al. (2020) sample which includes high Teq inflated Gas Giants with radii exceeding 14 R⊕ and masses as low as ~50 M⊕. For example, WASP-39 b with a mass of 89.3 M⊕ (Mancini et al. 2018) has a radius larger than predicted for CMF = 0 using the appropriate model from Fortney et al. (2007) and therefore should have a Jupiter composition. However, with a period of 4.06 days and Teq = 1166 K, this planet could be experiencing significant tidal inflation (Millholland et al. 2020) which may be consistent with a Saturn composition and an inflated super-Jupiter radius. It will be important to determine the reason that the Teq ≤ 600K Gas Giant sample does not have a mass overlap between Saturn and Jupiter composition planets in the mass range 50 – 120 M⊕.

**~4.4 Sub-Neptune Radius Gap**

Fulton et al. (2017) identified a gap in the radius distribution of small planets between 1.5 and 2.0 R⊕. Possible mechanisms considered for this radius gap include photoevaporation, gas poor formation environment, impact erosion, and core powered mass loss (Fulton et al. 2017; Ginzburg et al. 2018; Misener and Schlichting 2021). Zeng et al. (2019) identified planet equilibrium temperature as the most important parameter correlated with the radius valley. With only 57 planets having radii from 1.0 to 4.5 R⊕, the sample of planets in this catalog is currently too small to provide statistically meaningful conclusions in relation to the radius gap identified by Fulton et al. (2017).

The radius distribution for the entire catalog is provided in Table 10. There is a slight drop in the population at the radius gap identified by Fulton et al. (2017). This slight population drop, as compared with the deeper gap found by Fulton et al. (2017), could be consistent with the photoevaporation explanation for the sub-Neptune radius valley since photoevaporation should have less impact at Teq ≤ 600 K than at higher temperatures (Lopez & Fortney 2013), and therefore should produce a deeper valley at higher equilibrium temperatures. However, the sample in the catalog is currently too small to consider the small radius valley population drop sufficiently characterized. Instead, the current state of the radius distribution in this catalog is simply noted.

**Table 10: Planet Radius distribution**

| Radius Range (R⊕) | n/93 Catalog | n/101 Catalog + Solar System |
|---|---|---|
| < 1.0 | 3 | 6 |
| 1.0 – 1.5 | 12 | 13 |
| 1.5 – 2.0 | 10 | 10 |
| 2.0 – 2.5 | 13 | 13 |
| 2.5 – 3.0 | 11 | 11 |
| 3.0 – 4.5 | 11 | 13 |
| 4.5 – 7.0 | 3 | 3 |
| 7.0 – 10.0 | 14 | 15 |
| 10.0 – 14.0 | 16 | 17 |

**~4.5 Sub-Saturn Mass-Radius Desert**

The catalog sample has a deep radius desert between 4.5 and 7.5 $R_⊕$ - a radius range that includes only 5 planets. Four of the five planets are gas-rich sub-Neptunes and gas-rich Neptunes with masses between 6 and 18 $M_⊕$ (Tables 5 and 10). One possible explanation for this ***super-Neptune radius desert*** in the framework of the core accretion scenario is that the planets in the desert represent the small fraction of planets that accumulate a $f_{H-He}$ approaching the crossover gas fraction needed to initiate runaway accretion at approximately the same time the gas disk becomes significantly depleted or dissipates. This explanation is consistent with most planets in this radius range having a mass below pebble isolation mass and predicted mass for runaway accretion (~20 $M_⊕$) instead of a larger mass between the masses of Neptune and Saturn.

The sample also appears to have a sub-Saturn mass desert predicted by core accretion models (Ida & Lin 2004; Bertaux & Ivanova 2022) as there are only 6 planets in the sample with a mass between 30 and 100 $M_⊕$. It is interesting to note that most of these planets in the sub-Saturn mass desert have radii either smaller (one – 3.89 $R_⊕$) or larger (four – 7.9 – 8.9 $R_⊕$) than the sample of super-Neptune radius desert planets with radii from 4.5 – 7.5 $R_⊕$.

All planets in the $T_{eq} ≤ 600$ K catalog with a mass <180 $M_⊕$ are plotted in Figure 2 which reveals that the planets in the super-Neptune radius desert and the sub-Saturn mass desert in combination border a nearly barren mass-radius desert. This ***sub-Saturn mass-radius desert*** can be identified as the boxed region in Figure 2 with only one planet that has both a mass > 20 $M_⊕$ and a radius in the range 4.0 – 7.5 $R_⊕$. This characteristic of the $T_{eq} ≤ 600K$ sample is in contrast to the sample of Otegi et al. (2020) which has 12 planets with equilibrium temperature higher than 800 K that occupy the sub-Saturn mass-radius desert up to a mass of 70 $M_⊕$ (Figure 3).

The existence of a *sub-Saturn mass-radius* desert in the $T_{eq} ≤ 600K$ sample appears to be consistent with predictions of the core accretion scenario in several ways. First the lower mass limit of the desert, 20 $M_⊕$, is consistent with the pebble isolation mass (Lambrechts et al. 2014) and the critical core mass for runaway accretion (Pollack et al. 1996; Ida & Lin 2004) and would suggest that, for planets with $T_{eq} ≤ 600K$, most planets that reach the critical core mass of 20 $M_⊕$ do experience runaway accretion and rapidly grow to a mass exceeding ~100 $M_⊕$. Second, the upper radius limit of the desert is ~7.5 $R_⊕$ and therefore planets the planets forming the upper edge of the desert have H-He gas fractions exceeding 50%. This is consistent with runaway gas accretion for planets that exceed 20 $M_⊕$ resulting in planets with a Gas Giant composition.

The sample of Otegi et al. (2020) illustrates that the mostly barren state of the sub-Saturn Mass-Radius desert appears to be a characteristic specific to the $T_{eq} ≤ 600$ K sample and does not persist for samples with $T_{eq} > 800$ K (Figure 3). A thorough examination of the reasons for this difference is beyond the scope of this paper, but possible explanations include formation environment, gas disk properties, migration history, tidal inflationary effects (e.g. Millholland et al. 2020), or stellar irradiation effects. The orbital periods for the sample of planets from Otegi et al. (2020) found in the sub-Saturn mass-radius desert are all less than 13 days whereas most of the planets in the $T_{eq} ≤ 600$ K sample that border the desert edges have orbital periods exceeding 35 days (Tables 5 & 6).

**~4.6 Gas Giants orbiting M-dwarf stars**

Laughlin et al. (2004) predicted that Jupiter mass planets orbiting M-dwarf stars should be rare. This prediction may be consistent with the sample of Gas Giants as only 4/27 Gas Giants in the catalog (TOI-530 b, TOI-1899 b, TOI-3235 b, and TOI-3984 A b) orbit M-dwarfs. The most massive of these Gas Giant composition planets is TOI-1899 b with a mass of 0.67 Jupiter masses (212.9 $M_⊕$). Therefore, the $T_{eq} ≤ 600$ K sample does not contain any Jupiter mass planets orbiting M-dwarfs consistent with the prediction of Laughlin et al. (2004). Gan et al. (2022) noted that Gas Giants orbiting M-dwarfs all orbit high metallicity stars as predicted by the core accretion theory.

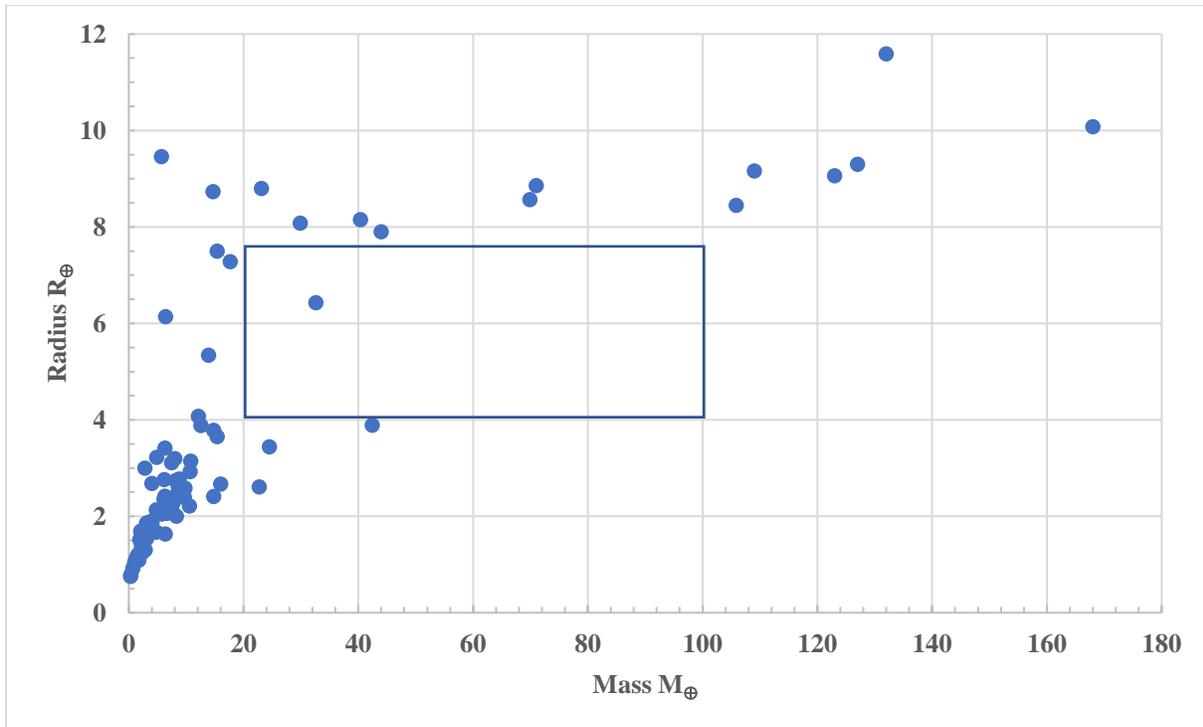

**Figure 2:** Mass-Radius plot for all planets (blue dots) in the catalog with a mass < 180 $M_\oplus$. The blue box outlines the sub-Saturn mass-radius desert found in this sample of $T_{eq} \leq 600K$ planets.

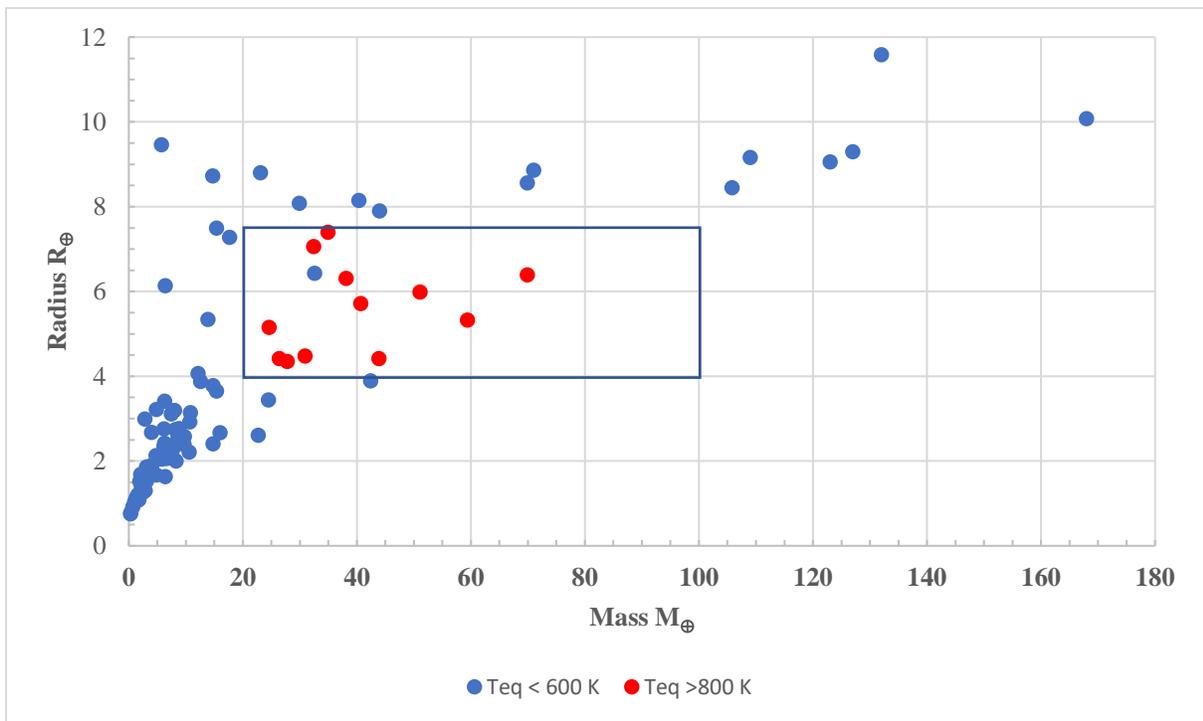

Figure 3: Mass-Radius plot for all planets (blue dots) in the catalog with a mass < 180 $M_\oplus$. The blue box outlines the sub-Saturn mass-radius desert found in this sample of $T_{eq} \leq 600K$ planets. The red dots in Figure 3 are planets with $T_{eq} > 800$ K from the sample of Otegi et al. (2020) that are located in the sub-Saturn mass-radius desert illustrating that the mostly barren mass-radius desert is a characteristic of planet samples with $T_{eq} \leq 600K$.

**~4.7 Critical Core Mass for Gas Accretion**

The low equilibrium temperatures for the planets in this catalog may provide an opportunity to identify the critical core mass for gas accretion since the post formation evolution of the Terrestrial and sub-Neptune planets in this sample should be less impacted by photoevaporation than higher Teq samples (Lopez & Fortney 2013). Ida & Lin (2004) identified this critical mass as ~4 $M_\oplus$.

Recently Modirrousta-Galian & Korenaga (2023) investigated three regimes of atmospheric erosion for super-Earth and sub-Neptune planets. They noted that for very small mass planets the probability of hosting a primordial H-He atmosphere is close to zero and at some mass the probability approaches one. This implies that there must be a critical mass between these limits within which approximately half of all planets have a primordial atmosphere. This critical mass is in the range 3 – 8 $M_\oplus$ (Modirrousta-Galian & Korenaga 2023). The sample of planets in this catalog is examined to see if this mass range can be further constrained since any low mass planets that acquired a primordial H-He envelope are more likely to maintain it for longer periods than planets exposed to higher stellar flux (Lopez & Fortney (2013).

The planets in this catalog with masses < 18 $M_\oplus$ were sorted into 6 composition classes: Terrestrial (T), gas-rich Terrestrial (grT), gas poor Neptune (gpN), Neptune (N), gas-rich Neptune (grN), and Saturn (S) for the mass ranges < 2.75 $M_\oplus$, 2.75 – 4.75 $M_\oplus$, 4.75 – 6.75 $M_\oplus$, 6.75 – 10.75 $M_\oplus$, and 10.75 – 17.75 $M_\oplus$. The results of this sorting are summarized in Table 11.

The population composition distribution in Table 11 suggests that the mass range 4.75 – 6.75 $M_\oplus$ appears to be the critical mass range where approximately half of the planets are Rock-Ice Giants that have accreted a substantial H-He envelope (1.0 – 49.9 % H-He by mass) and approximately half of the planets are Terrestrial (<1.0 % H-He by mass). This mass range has 5 Terrestrial planets and 7 Rock-Ice Giants and also has at least one planet for 5 of the 6 composition classes that occur for planets in this sample with a mass less than 18 $M_\oplus$ (Table 11).

The mass-composition distribution in Table 11 also reveals an increasing gas fraction among the sample population with increasing mass. Whereas the range 4.75 – 6.75 $M_\oplus$ has representatives for all Terrestrial and Neptune composition classes, 10/15 planets in the range 6.75 – 10.75 $M_\oplus$ are gas-poor Neptune composition. In the mass range 10.75 – 17.75 $M_\oplus$ 8/11 planets are Neptune, gas-rich Neptune, and Saturn composition whereas only one is a gas poor Neptune. Below 4.75 $M_\oplus$, there are no Neptune or gas-rich Neptune composition planets. Taken as a whole these trends add support to the conclusion that the mass range 4.75 – 6.75 $M_\oplus$ is the critical core mass range within which accumulation of a substantial H-He envelope begins.

**~4.8 Radius as a Proxy for Composition**

One of the important aspects of the sample of planets in this catalog is the strong connection between radius and composition for planets with Teq ≤ 600K (Table 8). Among the 93 planets in the catalog sample only four, or 4.3% of the sample, do not match the radius as a proxy for composition trends summarized in Table 8. The four planets with a composition not predicted from the radius are LHS 1140 b, K2-263 b, GJ 143 b, and Kepler 51 d. Each of these planets has been discussed in sections 3 and 4. The first 3 planets could be examples of planets that formed in a gas depleted environment that prevented accumulation of the amount of gas expected for the core mass (e.g. Lee & Chiang 2016) or were stripped of some or all of the accumulated H-He envelope by giant impacts (e.g. Roy et al. 2022). Kepler 51 d can be explained as resulting from a young age of 500 Myr. Libby-Roberts et al. (2020) found that at an age of 5 Gyr Kepler 51 d will have a mass and radius consistent with the trends in Table 8.

The strong correlation between radius and composition for the planets with Teq ≤ 600 K suggests that, for planets meeting the Teq cutoff and radius uncertainty standard of this catalog, but with no measured mass or a mass measurement that does not meet the uncertainty standard, radius can serve as a proxy for the planetary composition class as indicated in Table 8.

**Table 11: Composition Distribution for Identifying Critical Core Mass for H-He Accretion**

| Mass Range M⊕ | T | grT | gpN | N | grN | S |
|---|---|---|---|---|---|---|
| < 2.75 | 14 | 3 | 0 | 0 | 0 | 0 |
| 2.75 – 4.75 | 1 | 7 | 2 | 0 | 0 | 0 |
| 4.75 – 6.75 | 1 | 4 | 3 | 2 | 2 | 0 |
| 6.75 – 10.75 | 0 | 3 | 10 | 3 | 0 | 0 |
| 10.75 – 17.75 | 0 | 1 | 1 | 5 | 3 | 1 |

Composition codes: Terrestrial (T), gas-rich Terrestrial (grT), gas poor Neptune (gpN), Neptune (N), gas-rich Neptune (grN), and Saturn (S)

~5. Conclusions

In this paper, a catalog of planets with radius uncertainty ≤ 8%, mass uncertainty ≤ 27%, and $T_{eq}$ ≤ 600 K was identified from a search of the NASA Exoplanet Archive. The planets in this catalog should provide closer analogs to the planets of the Solar System, better candidates for habitable environments than higher equilibrium temperature exoplanet samples, and can be compared against predictions of planetary formation models. The main characteristics of the planets in this sample are:

~(1) Terrestrial composition planets have masses < 3.0 M⊕ and radii < 1.4 R⊕ whereas sub-Neptune composition planets have masses > 3.0 M⊕ and radii > 2.25 R⊕.

~(2) Planets with radii in the range 1.5 – 2.25 R⊕ are not sub-Neptunes, but are consistent with a gas-rich Terrestrial composition with $f_{H-He}$ from 0.1 – 0.5% by mass or a supercritical hydrosphere Terrestrial composition with $f_{H2O}$ from 5 – 20% by mass.

~(3) Rock-Ice Giants can be broken into gas-poor Neptune ($f_{H-He}$ 1 – 4 % by mass), Neptune ($f_{H-He}$ 4 – 20% by mass), and gas-rich Neptune ($f_{H-He}$ 20 – 49% by mass). Despite the large range in $f_{H-He}$ (~2 – 40 % by mass) and radius (2.3 – 7.5 R⊕) found in the catalog Rock-Ice Giant sample, only 3/35 Rock-Ice Giant composition planets have a mass exceeding 18 M⊕.

~(4) The sample population has a deep *super-Neptune radius desert* between 4.5 and 7.5 R⊕ and a sub-Saturn mass desert between 30 and 100 M⊕. The population of planets found in these deserts outline the edges of a mostly barren ***sub-Saturn mass-radius desert*** which is defined by the parameter space for masses > 20 M⊕ and in the radius range 4.0 – 7.5 R⊕ (Figure 2). This *sub-Saturn mass-radius desert* appears to be consistent with predictions of the core accretion scenario but an explanation will be needed as to why this desert is occupied by numerous planets with equilibrium temperatures > 800 K (Figure 3).

~(5) Gas Giants can be divided into Saturns (M⊕ <130, R⊕ <10.1, $f_{H-He}$ 50.0 – 80.0 % by mass) and Jupiters (M⊕ >130 and R⊕ >10.1, $f_{H-He}$ 80.1 – 100.0 % by mass) with no mass overlap between Saturn and Jupiter composition planets at masses <130 M⊕ and >180 M⊕.

~(6) Several mini-Saturn composition planets are found overlapping the mass range for Neptune composition planets from 15 – 30 M⊕.

~(7) Based upon the mass-composition distribution of planets in this sample, the critical core mass needed to begin accumulating a substantial H-He envelope appears to fall in the range 4.75 – 6.75 M⊕.

~(8) For over 95% of the planets in this $T_{eq}$ ≤ 600 K sample, radius serves as a direct proxy for planetary composition (Table 8) covering radii from < 1.0 R⊕ to 13.5 R⊕.

Many of the characteristics of this sample noted above appear to be consistent with the predictions of the core accretion scenario for planet formation including predictions related to the critical core mass for gas accretion (>~ 4 M⊕), critical mass for runaway gas accretion and the pebble isolation mass (~20 M⊕), and the existence of a sub-Saturn mass desert from 30 – 100 M⊕.

Finally, this catalog is presented as version 1.0 and the tables will be updated as new planets are identified meeting the catalog selection criteria, or as planets not currently meeting the uncertainty requirements are updated with mass and radius measurements that meet the selection criteria of this catalog. Updated tables for this catalog may be obtained by contacting the author.

**Acknowledgements**

This research has made use of the NASA Exoplanet Archive, which is operated by the California Institute of Technology, under contract with the National Aeronautics and Space Administration under the Exoplanet Exploration Program. This research has made use of NASA's Astrophysics Data System Bibliographic Services.